\documentclass[twocolumn,english,amsmath,ammsymb]{revtex4-1}
\usepackage[T1]{fontenc}
\usepackage[latin9]{inputenc}
\usepackage{color}
\usepackage{amsmath}
\usepackage{amssymb}
\usepackage{graphicx}
\usepackage{babel}
\begin{document}
\title{Emergent Mott-insulators at non-integer fillings and devil's staircase
induced by attractive interaction in many-body polarons}
\author{Jian-Hua Zeng}
\affiliation{Institute for Theoretical Physics, SPTE, South China Normal University,
Guangzhou 510006, China}
\affiliation{Guangdong Provincial Key Laboratory of Quantum Engineering and Quantum
Materials, Guangzhou 510006, China}
\author{Su Yi}
\email{syi@itp.ac.cn}

\affiliation{CAS Key Laboratory of Theoretical Physics, Institute of Theoretical
Physics, Chinese Academy of Sciences, Beijing 100190, China }
\affiliation{School of Physical Sciences and CAS Center for Excellence in Topological
Quantum Computation, University of Chinese Academy of Sciences, Beijing
100049, China}
\affiliation{Peng Huanwu Collaborative Center for Research and Education, Beihang
University, Beijing 100191, China}
\author{Liang He}
\email{liang.he@scnu.edu.cn}

\affiliation{Institute for Theoretical Physics, SPTE, South China Normal University,
Guangzhou 510006, China}
\affiliation{Guangdong Provincial Key Laboratory of Quantum Engineering and Quantum
Materials, Guangdong-Hong Kong Joint Laboratory of Quantum Matter,
South China Normal University, Guangzhou 510006, China}
\begin{abstract}
We investigate the ground state properties of an ultracold atom system
consisting of many-body polarons, quasiparticles formed by impurity
atoms in optical lattices immersing in a Bose-Einstein condensate.
We find the nearest-neighbor attractive interaction between polarons
can give rise to rich physics that is peculiar to this system. In
a relatively shallow optical lattice, the attractive interaction can
drive the system being in a self-bound superfluid phase with its particle
density distribution manifesting a self-concentrated structure. While
in a relatively deep optical lattice, the attractive interaction can
drive the system forming the Mott-insulator phase even though the
global filling factor is not integer. Interestingly, in the Mott-insulator
regime, the system can support a series of different Mott-insulators
with their effective density manifesting a devil's staircase structure
with respect to the strength of attractive interaction. Detailed estimation
on relevant experimental parameters shows that these rich physics
can be readily observed in current experimental setups.
\end{abstract}
\maketitle

\section{Introduction}

Since the concept of polaron was first proposed by Landau and Pekar
in their study of moving electrons in dielectric crystals \citep{Zh_Eksp_Teor_Fiz.Landau.1948.polaron},
understanding the properties of impurities interacting with quantum
baths has been an important research field in condensed matter physics.
This is mainly due to the fact that polarons can play an important
role in understanding properties of various important condensed matter
systems such as high-$T_{c}$ superconductors \citep{Nature.Lanzara.2001.highTsc,RevModPhys.Lee.2006.highTsc}
and semiconductors \citep{RevModPhys.Gershenson.2006.semiconductors,OrganicElectronics.Nianduan.2018.semiconductor},
also naturally attracting much research interest in the context of
quantum simulations with ultracold atomic gases, where there has been
continuous effort devoted to the investigation of polaron physics
in the last two decades \citep{PhysRevA.Astrakharchik.2004.fewbody,PhysRevLett.Kenneth.2006.fermipolaronevidence,PhysRevLett.Ospelkaus.2006.fermipolaronevidence,PhysRevA.Chevy.2006,PhysRevA.Kalas.2006.fewbody,PhysRevLett.Cucchietti.2006.fewbody,PhysRevB.Prokof'ev.2008.fermitheoretical,PhysRevLett.Schirotzek.2009.femionpolaron,PhysRevLett.Palzer.2009,PhysRevLett.Nascimbene.2009,PhysRevLett.Frank.2010.manybodypolaron,Nature.Kohstall2012,PhysRevLett.Zhang.2012.fermipolaron,PhysRevA.Casteels.2012.theoreticalEfforts,PhysRevA.Schmidt.2012.fermitheoretical,PhysRevA.Catani.2012,Nature.Physics.Fukuhara.2013,PhysRevA.Rath.2013.bosetheoretical,PhysRevA.Li.2014.theoreticalEffort,ScientificReports.Grusdt.2015,PhysRevLett.Christensen.2015,PhysRevA.Ardila.2015.bosepolaron,PhysRevLett.Nils.2016.bosepolaronexper,PhysRevLett.Hu.2016,Science.Marko.2016.manybodydynamic,PhysRevLett.Shchadilova.2016.bosetheoretical,NewJournalofPhysics.Bellotti.2016.fewbody,PhysRevX.Schmidt.2016.few,PhysRevLett.Scazza.2017.fermipolaron,RevModPhys.Greene.2017.fewbody,PhysRevA.VanLoon.2018.manybodypolaron,Nature.Skou.2021.manybody,PhysRevX.Dolgirev.2021}. 

In this context, single- and few-polaron systems have been largely
studied, and their rich physics has been revealed \citep{ReportsonProgressinPhysics.Schmidt.2018,ReportsonProgressinPhysics.Massignan.2014.fermipolaron,Bloch_Nat_Phys_2022}.
Particularly, the high tunability of optical lattices also motivates
the investigations of polarons in optical lattices, i.e., polarons
formed by impurity atoms in optical lattices immersed in a Bose-Einstein
condensate (BEC) \citep{PhysRevA.Bruderer.2007.Bogoliubov,NewJournalofPhysics.AlexanderKlein.2007,PhysRevA.Privitera.2010,PhysRevA.Yin.2015},
or coupled to another species of atoms in optical lattices \citep{JournalofPhysicsB.Yordanov.2023.R2P,arXiv.Colussi.2022.R2P,arXiv.ding.2022.R2P},
where interesting phenomena such as the clustering, self-trapping
effect, polaronic slowing, strong influences on polaronic properties
imposed by the bath near its phase transition, and formation of bipolarons
have been found.

However, despite these rich physics revealed in single- and few-polaron
systems \citep{ReportsonProgressinPhysics.Schmidt.2018,ReportsonProgressinPhysics.Massignan.2014.fermipolaron,Bloch_Nat_Phys_2022},
the physics associated with many-polaron systems has been far less
studied, although the existence of induced effective interactions
between polarons has been revealed \citep{PhysRevA.Bruderer.2007.Bogoliubov,NewJournalofPhysics.AlexanderKlein.2007,PhysRevA.Privitera.2010,NewJournalofPhysics.Santamore.2011.attractivePolaronInteraction,PhysRevA.Yin.2015}.
In fact, for polarons in optical lattices, these induced effective
interactions not only assume an on-site part but also assume an off-site
part. This reminds one of the rich many-body physics of the ultracold
gases with dipolar interactions in optical lattices (see, e.g., Ref.~\citep{ChemicalReviews.Baranov.2012.Dipolar}
and references therein). For instance, the so-called devil's staircase,
which was first identified in long-range interacting lattice models
of classical particles and spins \citep{PhysRevB.Hubbard.1978,PhysRevLett.Fisher.1980,PhysRevB.Bak.1980.devil,PhysRevLett.Bak.1982.devil},
later also found in other different systems, such as liquid crystals
\citep{RevModPhys.Takezoe.2010}, quantum models of dimers \citep{Fradkin_PRB_2004,Schlittler_PRL_2015},
spin-valve systems \citep{Matsuda_PRL_2015}, fractional quantum hall
systems \citep{Rotondo_PRL_2016}, is identified in ultracold gases
with dipolar interactions in optical lattices \citep{PhysRevB.Burnell.2009.devil,PhysRevLett.Capogrosso.2010,PhysRevA.Ohgoe.2012.devil,PhysRevA.Zhang.2021.dipolar.devil}. 

In these regards, one naturally expects that the induced effective
interactions in many-polaron systems could give rise to novel physics
that is absent in single- and few-polaron systems. Particularly, in
considerable cases, since the quantum bath that directly interacts
with impurity atoms consists of quantum harmonic oscillators, these
induced interactions between polarons are usually attractive \citep{PhysRevA.Bruderer.2007.Bogoliubov,NewJournalofPhysics.AlexanderKlein.2007,PhysRevA.Privitera.2010,NewJournalofPhysics.Santamore.2011.attractivePolaronInteraction,PhysRevA.Yin.2015}.
Noticing in addition that polarons in ultracold atomic systems also
inherit the repulsive contact interaction from the impurity atoms,
this thus naturally gives rise to the interesting question of physical
influences from the competition between these two types of interactions
in many-body polarons.

\begin{figure}[h]
\begin{centering}
\includegraphics[width=3in]{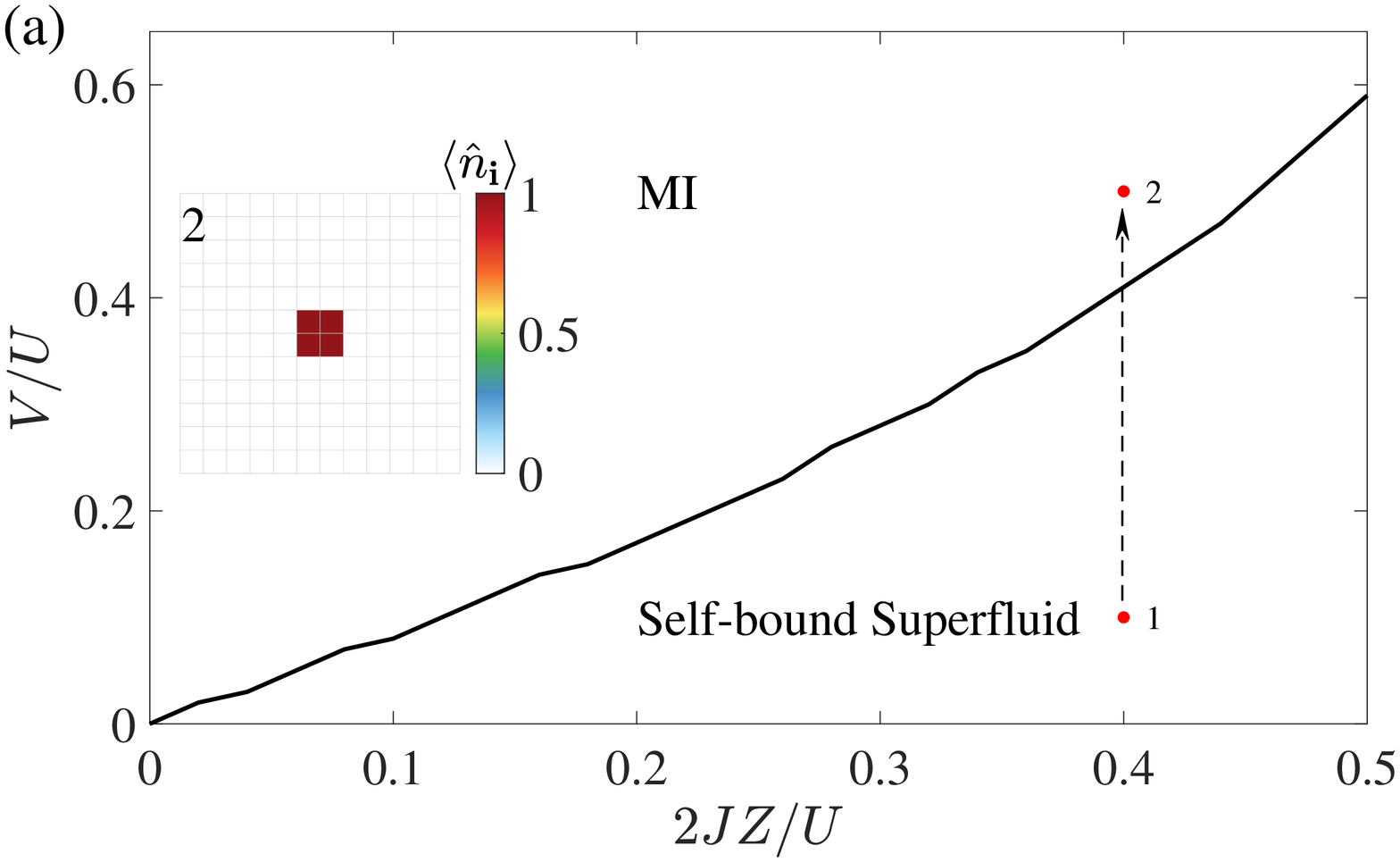}
\par\end{centering}
\begin{centering}
\includegraphics[width=3in]{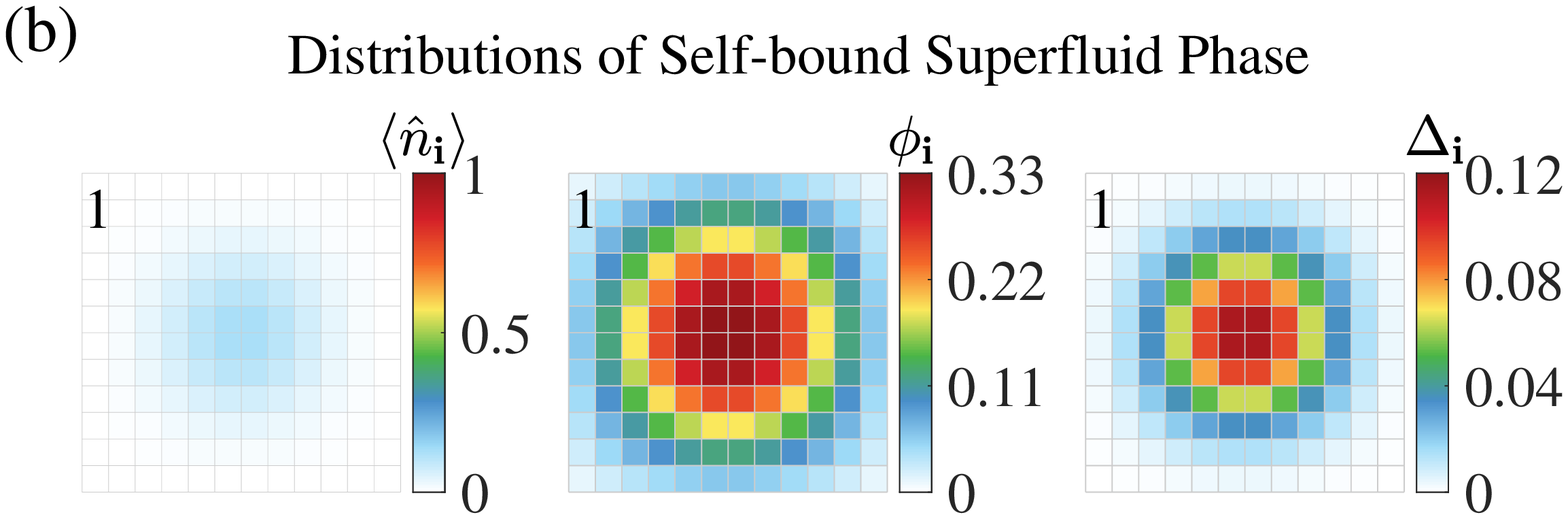}
\par\end{centering}
\caption{\label{fig:low_filling}Phase diagram and typical real-space distributions
of the system at the filling factor $\rho=1/36$ on a $12\times12$
square lattice. (a) Phase diagram at $\rho=1/36$, which features
a transition between the self-bound superfluid phase and the Mott-insulator
(MI) phase. Inset: Real-space polaron density $\langle\hat{n}_{\mathbf{i}}\rangle$
distribution of a MI {[}upper red dot in (a), $2JZ/U=0.4,V/U=0.5${]}.
(b) Typical $\langle\hat{n}_{\mathbf{i}}\rangle$, $\phi_{\mathbf{i}}$
and $\Delta_{\mathbf{i}}$ distributions of the self-bound superfluid
{[}lower red dot in (a), $2JZ/U=0.4,V/U=0.1${]}, featuring a self-concentrated
structure. See text for more details. }
\end{figure}

Motivated by recent ultracold atom experiments, here we address this
question by investigating a system where one species of bosonic atoms
(impurities) trapped in an optical lattice is immersed in a BEC formed
by another species of atoms. The interaction between the impurity
atoms and the Bogoliubov (phonon) modes of the BEC drives the formation
of polarons, and gives rise to an interacting many-body polaron system.
We investigate the physical influences of the attractive interaction
between polarons by establishing the phase diagrams of the system
at different filling factors (see Fig.~\ref{fig:low_filling} and
Fig.~\ref{fig:intermediate_filling}) and find that the attractive
interaction can give rise to rich physics. More specifically, we find
the followings.

(i) Self-bound superfluid and emergent Mott-insulators (MI) at non-integer
filling factors. In a relatively shallow optical lattice, the attractive
interaction can drive the system being in a self-bound superfluid
phase with its particle density distribution manifesting a self-concentrated
structure {[}see Fig.~\ref{fig:low_filling}(b){]}. While in a relatively
deep optical lattice, the attractive interaction can drive the system
forming the Mott-insulator phase even though the global filling factor
is not integer {[}see the inset of Fig.~\ref{fig:low_filling}(a){]}. 

(ii) Reentrance to self-bound superfluid and devil\textquoteright s
staircase induced by attractive interaction. At intermediate filling
factors and in the relative small hopping regime, increasing the attractive
interaction strength can first drive the system from self-bound superfluid
phase into Mott-insulator phase, then back to self-bound superfluid
phase again {[}see the vertical arrows and insets in Figs.~\ref{fig:intermediate_filling}(a,
c){]}. In fact, a series of this type of reentrance can be present
in the system as long as the filling factor is large enough {[}see
for instance the vertical arrow and inset in Fig.~\ref{fig:intermediate_filling}(c)
with the corresponding filling factor $\rho=1/3${]}. Interestingly,
the system can also support a succession of incompressible Mott-insulator
states, dense in the parameter space (see Fig.~\ref{fig:devil_staircase}),
which is reminiscent of the devil's staircase in long-range interacting
system tuned by the filling factor, but here in our case, is driven
by an essentially short-range attractive interaction with the filling
factor of the system kept fixed. 

\section{System and model}

Motivated by related experiments, we consider the system where one
species of bosonic atoms (impurities) trapped in a square optical
lattice is immersed in a BEC formed by another species of atoms \citep{PhysRevA.Bruderer.2007.Bogoliubov,NewJournalofPhysics.Bruderer.2008.Bogoliubov,PhysRevA.Yin.2015}.
The system can be described by a Hamiltonian $\hat{H}_{\mathrm{sys}}$
consisting of three parts, i.e., $\hat{H}_{\mathrm{sys}}=\hat{H}_{I}+\hat{H}_{B}+\hat{H}_{\mathrm{int}}$.
Here, $\hat{H}_{I}$ is the Hamiltonian of impurity atoms assuming
the form of a conventional Bose Hubbard model, i.e., $\hat{H}_{I}=-\sum_{\langle\mathbf{i},\mathbf{j}\rangle}J_{0}\hat{a}_{\mathbf{i}}^{\dagger}\hat{a}_{\mathbf{j}}-\sum_{\mathbf{i}}\mu_{0}\hat{a}_{\mathbf{i}}^{\dagger}\hat{a}_{\mathbf{i}}+\sum_{\mathbf{i}}(U_{0}/2)\hat{a}_{\mathbf{i}}^{\dagger}\hat{a}_{\mathbf{i}}^{\dagger}\hat{a}_{\mathbf{i}}\hat{a}_{\mathbf{i}},$
with $\hat{a}_{\mathbf{i}}^{\dagger}(\hat{a}_{\mathbf{i}})$ being
the impurity creation (annihilation) operator at site $\mathbf{i}$
in the Wannier basis. The BEC is treated as a Bogoliubov phonon bath
described by $\hat{H}_{B}=\sum_{\mathbf{q}}\hbar\omega_{\mathbf{q}}\hat{\beta}_{\mathbf{q}}^{\dagger}\hat{\beta}_{\mathbf{q}}$,
where $\omega_{\mathbf{q}}$ is the Bogoliubov phonon spectrum with
momenta $\mathbf{q}$ and $\hat{\beta}_{\mathbf{q}}^{\dagger}(\hat{\beta}_{\mathbf{q}})$
is the creation (annihilation) operator for the Bogoliubov phonons.
The interaction between the impurities and phonons are described by
the Hamiltonian $\hat{H}_{\mathrm{int}}=\sum_{\mathbf{i}}\sum_{\mathrm{\mathbf{q}}}\hbar\omega_{\mathbf{q}}M_{\mathrm{\mathbf{q}}}e^{i\mathbf{q}\cdot\mathbf{r}_{\mathbf{i}}}(\hat{\beta}_{\mathbf{q}}+\hat{\beta}_{-\mathrm{\mathbf{q}}}^{\dagger})\hat{a}_{\mathbf{i}}^{\dagger}\hat{a}_{\mathbf{i}}+\text{h.c.}$,
where $M_{\mathrm{\mathbf{q}}}$ describes the impurity-phonon coupling
(see Appendix \ref{App:Exp_Par_Est} for details). 

Due to the interactions between the impurity atoms and the Bogoliubov
(phonon) modes of the BEC, the impurities and the phonons can form
quasi-particles, i.e., polarons \citep{PhysRevA.Bruderer.2007.Bogoliubov,NewJournalofPhysics.Bruderer.2008.Bogoliubov,PhysRevA.Yin.2015}.
Using Lang-Firsov polaron transformation \citep{PhysRevA.Bruderer.2007.Bogoliubov,NewJournalofPhysics.Bruderer.2008.Bogoliubov,PhysRevB.Maier.2011.Lang_Firsov,PhysRevA.Yin.2015},
which takes the form $\widetilde{H}\equiv e^{\hat{S}}\hat{H}_{\mathrm{sys}}e^{-\hat{S}}$
with $\hat{S}\equiv\sum_{\mathbf{i\mathrm{,}q}}\lambda_{\mathbf{q}}M_{\mathbf{q}}e^{i\mathbf{q}\cdot\mathbf{r}_{\mathbf{i}}}\left(\hat{\beta}_{-\mathbf{q}}^{\dagger}-\hat{\beta}_{\mathbf{q}}\right)\hat{a}_{\mathbf{i}}^{\dagger}\hat{a}_{\mathbf{i}}$,
where $\lambda_{\mathbf{q}}$ is the variational parameter of Lang-Firsov
polaron transformation to be determined self-consistently (see Appendix
\ref{App:Exp_Par_Est} for details), the transformed Hamiltonian $\widetilde{H}$
can be separated into a coherent part $\langle\widetilde{H}\rangle$
and an incoherent part. At low temperatures, the physics of the system
can be effectively described by the coherent part of the Hamiltonian
for the polarons after Lang-Firsov polaron transformation, since the
incoherent part is strongly suppressed in the low-temperature regime
\citep{PhysRevA.Bruderer.2007.Bogoliubov,NewJournalofPhysics.Bruderer.2008.Bogoliubov,PhysRevA.Yin.2015}.

Therefore, the effective Hamiltonian for the polarons reads (see Appendix
\ref{App:Exp_Par_Est} for more derivation details)
\begin{align}
\hat{H}= & -J\sum_{\langle\mathbf{i},\mathbf{j}\rangle}\hat{b}_{\mathbf{i}}^{\dagger}\hat{b}_{\mathbf{j}}+\frac{U}{2}\sum_{\mathbf{i}}\hat{n}_{\mathbf{i}}(\hat{n}_{\mathbf{i}}-1)-\sum_{\langle\mathbf{i},\mathbf{j}\rangle}\frac{V}{2}\hat{n}_{\mathbf{i}}\hat{n}_{\mathbf{j}},\label{eq:Hamiltonian}
\end{align}
where $\hat{b}_{\mathbf{i}}^{\dagger}$($\hat{b}_{\mathbf{i}}$) is
the creation (annihilation) operator of polarons at site $\mathbf{i}$
in the Wannier representation, $\hat{n}_{\mathbf{i}}\equiv\hat{b}_{\mathbf{i}}^{\dagger}\hat{b}_{\mathbf{i}}$
is the particle number operator that counts the number of polarons
on site $\mathbf{i}$, and $\langle\mathbf{i},\mathbf{j}\rangle$
denotes nearest neighbor lattice sites. Here, the Hamiltonian (\ref{eq:Hamiltonian})
assumes the form of the extended Bose-Hubbard model, where the first
two terms are the conventional hopping term with a hopping amplitude
$J$ and the on-site interaction term whose strength is specified
by $U$. The third term describes the induced nearest neighbor attractive
interaction \citep{PhysRevA.Bruderer.2007.Bogoliubov,NewJournalofPhysics.AlexanderKlein.2007,PhysRevA.Privitera.2010,NewJournalofPhysics.Santamore.2011.attractivePolaronInteraction,PhysRevA.Yin.2015}
between polarons whose strength is specified by $V$ ($V>0$). It
originates from the coupling between impurity atoms and the Bogoliubov
modes of the BEC \citep{PhysRevA.Bruderer.2007.Bogoliubov,NewJournalofPhysics.Bruderer.2008.Bogoliubov,PhysRevA.Yin.2015}. 

From the form of the Hamiltonian (\ref{eq:Hamiltonian}), we see that
its first two terms form the conventional Bose-Hubbard model \citep{PhysRevB.Fisher.1989}
which favor two homogeneous phases that respect the discrete translational
symmetry of the underlying lattice, i.e., homogeneous superfluid at
large $J/U$ and homogeneous Mott-insulators at integer filling factors
at small $J/U$. While for the nearest neighbor attractive interaction
term, it can drive the polarons to concentrate in space, making real-space
distributions of typical physical quantities, such as density distributions,
inhomogeneous. This thus breaks the discrete translational symmetry
of the system. In these regards, one would expect that the presence
of the nearest neighbor attractive interaction could give rise to
new physics beyond the one associated with conventional superfluid
to Mott-insulator transition. Indeed, as we shall see in the following
the attractive interaction can give rise to Mott-insulators at non-integer
fillings. Even more remarkably, it can drive the system to form a
series of incompressible ground states. This is reminiscent of the
devil's staircase in long-range interacting systems tuned by the filling
factor \citep{PhysRevB.Hubbard.1978,PhysRevLett.Fisher.1980,PhysRevLett.Bak.1982.devil,PhysRevLett.Capogrosso.2010},
but here in our case, is driven by the attractive interaction with
the filling factor of the system kept fixed.

\section{Results}

In the investigations to be presented in the following, we use the
bosonic Gutzwiller variational approach \citep{PhysRevB.Krauth.1992,PhysRevLett.Jaksch.1998,PhysRevB.Nicola.2012.Gutzwiller}
to investigate the ground state properties of the system, with the
variational ground state assuming the site-factorized form $|\mathrm{GW}\rangle=|\phi_{1}\rangle_{1}\otimes\ldots\otimes|\phi_{N_{\mathrm{lat}}}\rangle_{N_{\mathrm{lat}}}$.
Here, $N_{\mathrm{lat}}$ is the total number of the lattice sites
and $|\phi_{\mathbf{i}}\rangle_{\mathbf{i}}=\sum_{n=0}^{\infty}c_{n}^{(\mathbf{i})}|n\rangle_{\mathbf{i}}$
is the local wave function at site $\mathbf{i}$ with $|n\rangle_{\mathbf{i}}$
being the corresponding local occupation number state and $c_{n}^{(\mathbf{i})}$
being the variational parameter. The ground state is determined by
minimizing the total energy of the system within this variational
ansatz, i.e., $E(\{c_{n}^{(\mathbf{i})}\})=\langle\mathrm{GW}|\hat{H}|\mathrm{GW}\rangle$.
In the following, we investigate the ground state properties of the
system at different fixed filling factors $\rho\equiv N/N_{\mathrm{lat}}$,
with $N$ being the total number of polarons in the system. If not
specified in the text, a square lattice with the linear system size
$L=12$, the local occupation number cutoff $n_{\mathrm{max}}=13$,
and open boundary condition are chosen in the numerical results presented
in the following (periodic boundary condition can also be employed
and only gives rise to small differences). 

\subsection{Self-bound superfluid and emergent Mott-insulators at non-integer
fillings }

At low filling factor $\rho\equiv\langle\hat{N}\rangle/N_{\mathrm{lat}}$,
typical properties of the system are summarized in Fig.~\ref{fig:low_filling},
which shows a phase diagram of the system at the filling factor $\rho=1/36$
and typical real-space distributions of polaron density $\langle\hat{n}_{\mathbf{i}}\rangle$,
superfluid order parameter $\phi_{\mathbf{i}}\equiv\langle\hat{b}_{\mathbf{i}}\rangle$
and local density fluctuation $\Delta_{\mathbf{i}}\equiv\langle\hat{n}_{\mathbf{i}}^{2}\rangle-\langle\hat{n}_{\mathbf{i}}\rangle^{2}$.
In the large hopping amplitude regime {[}see lower right part of Fig.~\ref{fig:low_filling}(a){]},
the system breaks the $\mathrm{U(1)}$ symmetry and is in a superfluid
phase characterized by the existence of non-zero superfluid order
parameter $\phi_{\mathbf{i}}$. Interestingly, as one can notice from
Fig.~\ref{fig:low_filling}(b), the polaron density and superfluid
order parameter distributions peak at the center of the lattice. We
remark here that no external trapping potential is present in our
calculation, this real-space concentration reflects the influences
of the attractive interaction between polarons. In the following,
we thus refer to it as the self-bound superfluid phase. 

\begin{figure}[p]
\begin{centering}
\includegraphics[width=2.7in]{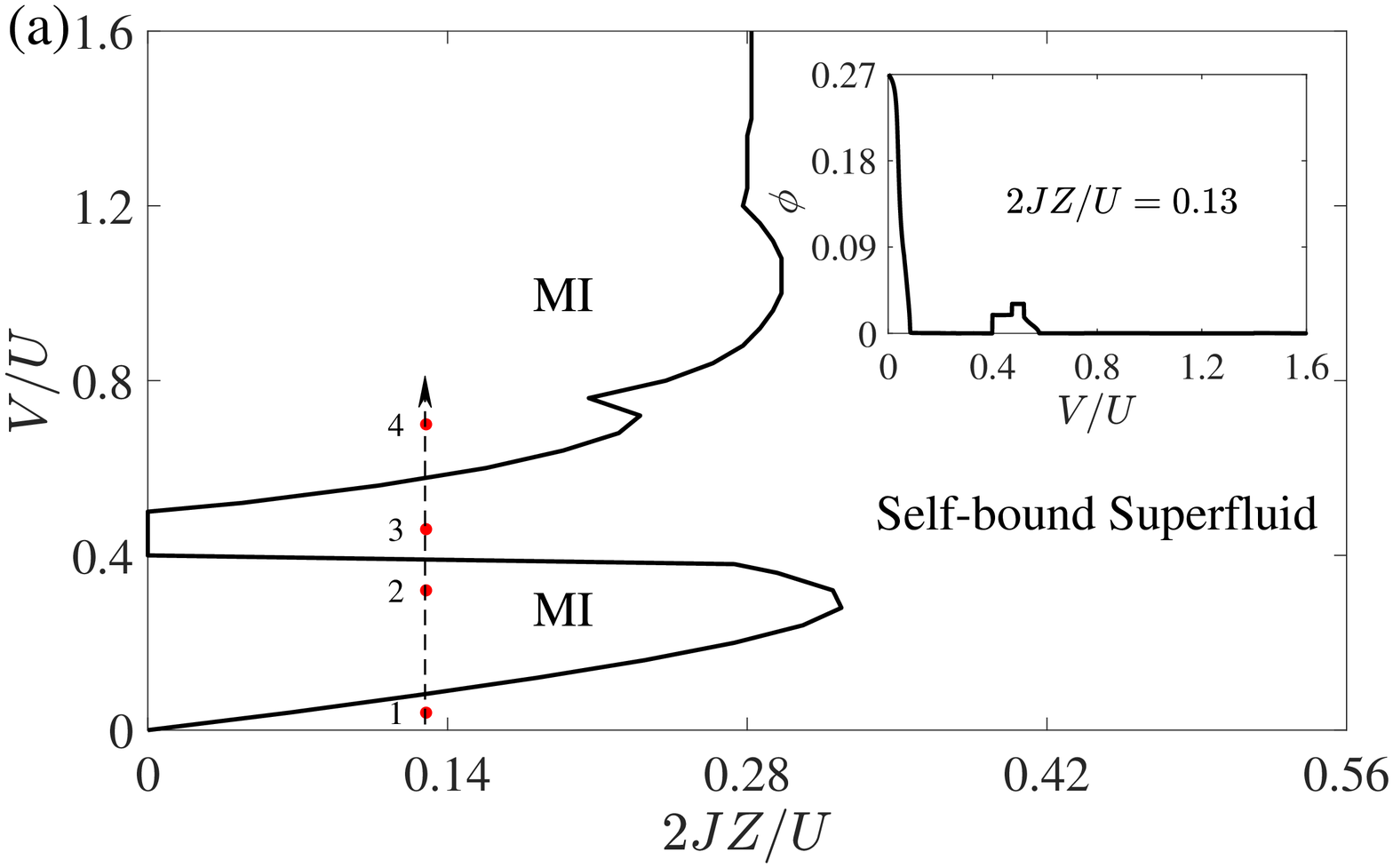}
\par\end{centering}
\begin{centering}
\includegraphics[width=2.7in]{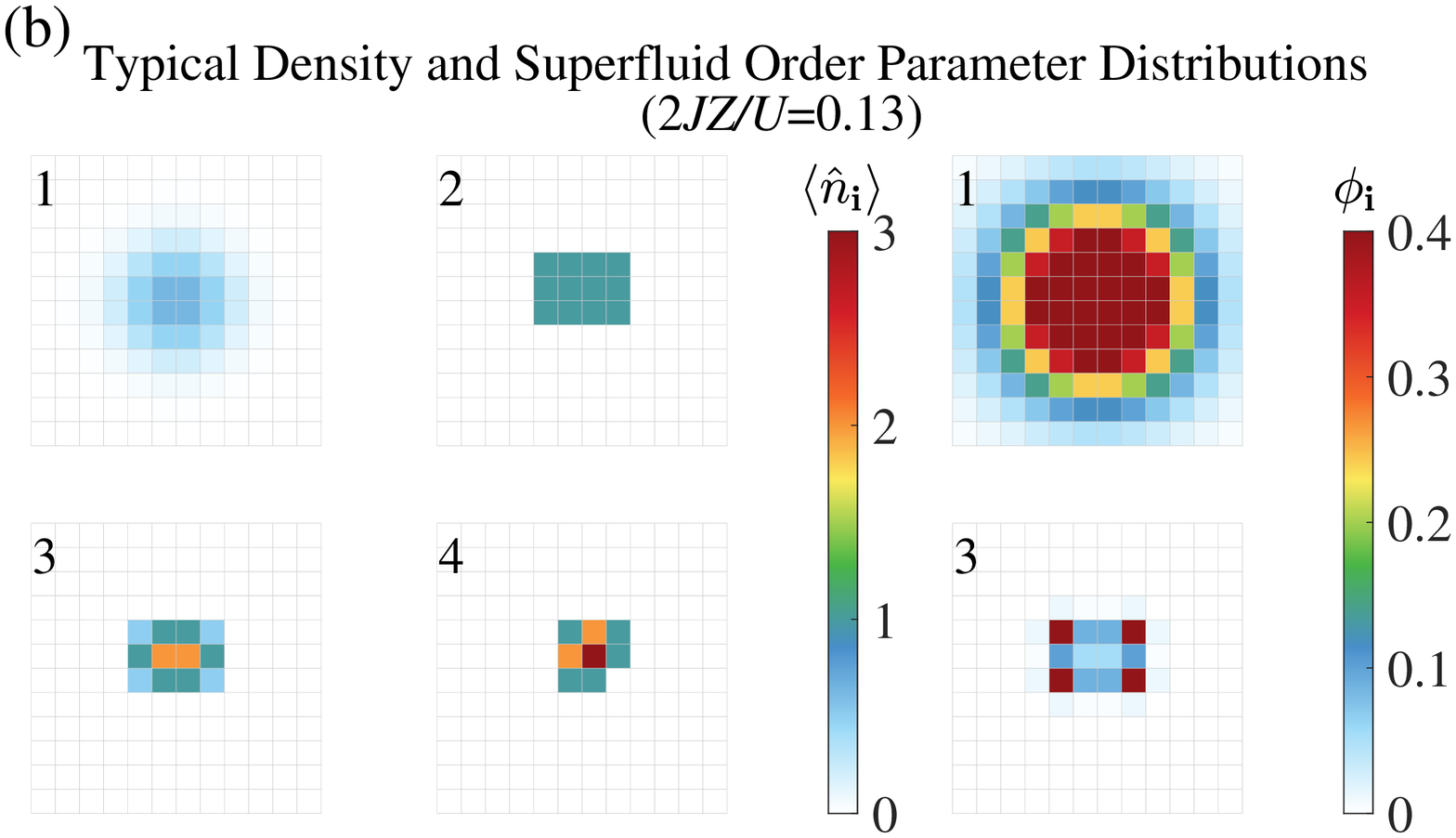}
\par\end{centering}
\begin{centering}
\includegraphics[width=2.7in]{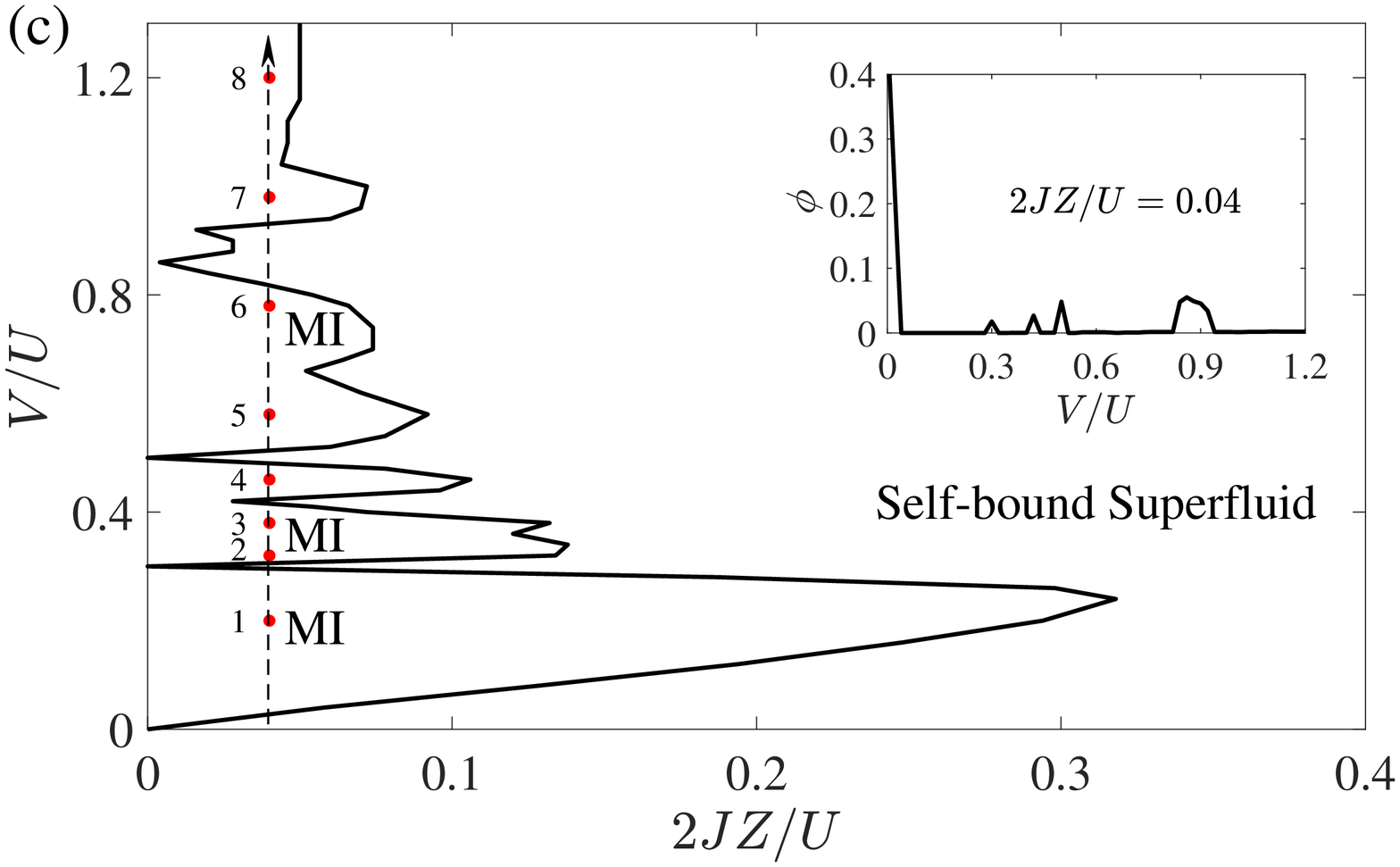}
\par\end{centering}
\begin{centering}
\includegraphics[width=2.7in]{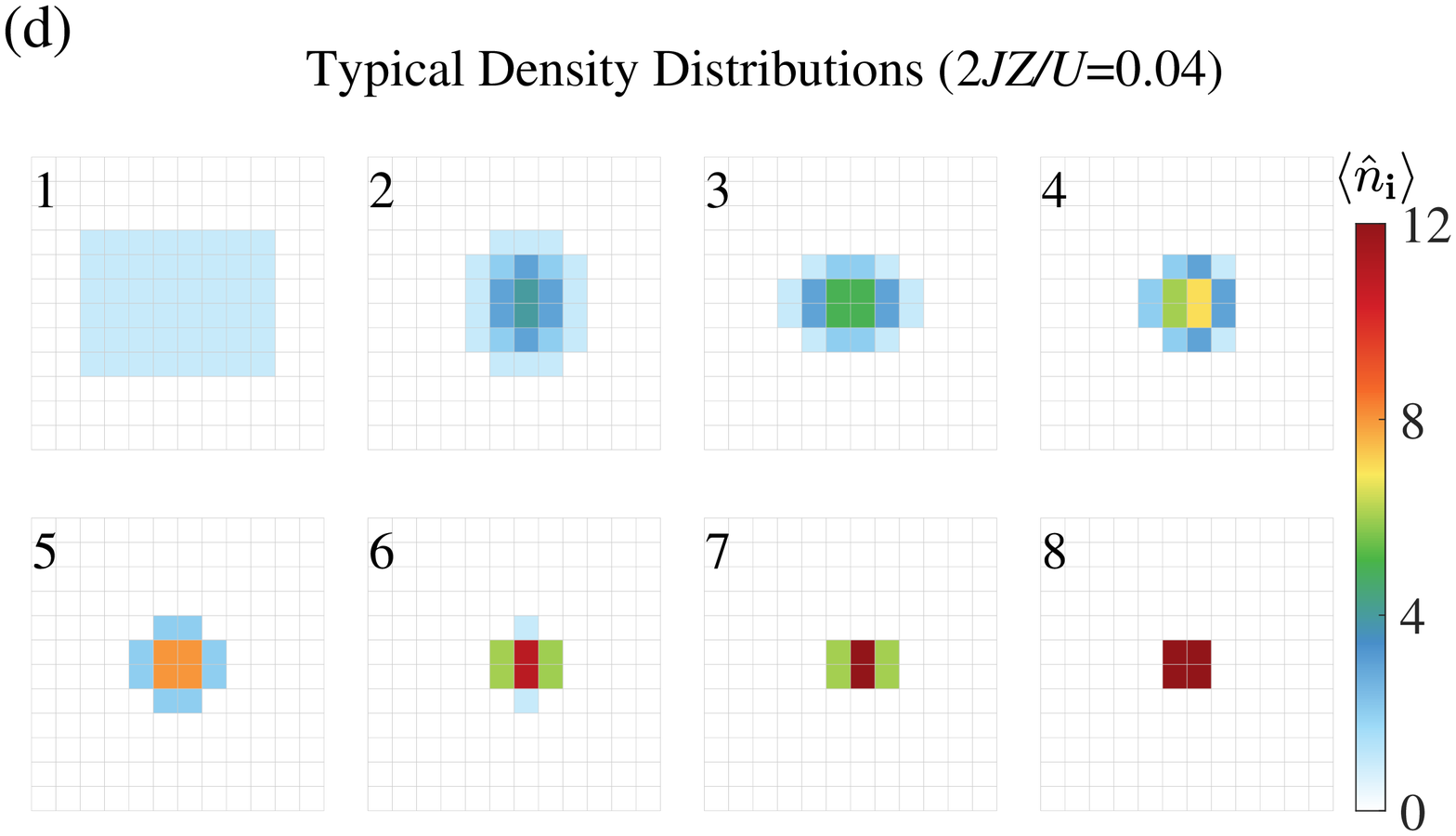}
\par\end{centering}
\caption{\label{fig:intermediate_filling}Phase diagrams and typical real-space
distributions at intermediate fillings. (a) Phase diagram at $\rho=1/12$.
The inset shows the $V/U$ dependence of the average superfluid order
parameter $\phi\equiv N_{\mathrm{lat}}^{-1}\sum_{\mathbf{i}}\phi_{\mathbf{i}}$
at a fixed hopping amplitude ($2JZ/U=0.13$, see also the arrow in
the main plot). (b) Typical density and superfluid order parameter
distributions that correspond to the red dots in (a). The values of
$V/U$ for the red dots marked by 1 to 4 are $0.04,\,0.32,\,0.46$,
and $0.7$, respectively. The $2JZ/U$ values for these dots are the
same, with $2JZ/U=0.13$. (c) Phase diagram at $\rho=1/3$. The inset
shows the $V/U$ dependence of $\phi$ at a fixed hopping amplitude
($2JZ/U=0.04$, see also the arrow in the main plot). (d) Density
distributions that correspond to the red dots in (c). The values of
$V/U$ for the red dots marked by 1 to 8 are $0.2$, $0.32$, $0.38$,
$0.46$, $0.58$, $0.78$, $0.98$, and $1.2$, respectively. The
$2JZ/U$ values for these dots are the same, with $2JZ/U=0.04$. See
text for more details. }
\end{figure}

Comparing to conventional homogeneous superfluid phases in similar
systems without attractive interactions, the spatial polaron density
and superfluid order parameter distributions of the self-bound superfluid
phase show that the attractive interaction can drive the polarons
to the central region of the system. This indicates although the filling
factor of the system in this case is well below unit filling, a strong
enough attractive interaction can still drive the emergence of a local
Mott-insulator phase by increasing the local filling factor or density
in the central region of the system. Indeed, as one can see from Fig.~\ref{fig:low_filling}(a),
when the attractive interaction strength is relatively strong compared
with the hopping amplitude, the system always forms a Mott-insulator
with vanishing superfluid order parameter and local density fluctuations.
Finally, we remark that because the phase diagram presented in Fig.~\ref{fig:low_filling}
and the ones to be presented in Fig.~\ref{fig:intermediate_filling}
are obtained within bosonic Gutzwiller variational approach, possible
strong quantum fluctuations that exist in the vicinity of the phase
boundaries are not well accounted within this mean-field approach.
These quantum fluctuations are expected to impose corrections on the
phase boundaries and local density fluctuations in these quantum critical
regimes. Although beyond the scope of the current work, it is interesting
to further investigate the influences from quantum fluctuations in
these quantum critical regimes by employing methods beyond mean-field,
such as the quantum Monte Carlo method \citep{PhysRevB.Capogrosso.2007.R2G,PhysRevA.Guglielmino.2010.R2G}
that has been applied to the study of Bose-Hubbard type models, the
quantum Gutzwiller approach developed recently \citep{PhysRevResearch.Caleffi.2020.R2G,SciPostPhys.Victor.2022.R2G},
etc. 

\subsection{Reentrance to self-bound superfluid and devil's staircase induced
by attractive interaction }

Noticing that in the above low filling case, the total particle number
of the system is quite small ($N=4$) and strongly restricts the number
of possible configurations of density distributions, therefore, one
naturally expects that at intermediate filling factors, the system
could manifest richer physics induced by the attractive interaction.
This motivates us to investigate the properties of the system at intermediate
filling factors, the results of which are summarized in Fig.~\ref{fig:intermediate_filling},
where two phase diagrams of the system at two different intermediate
filling factors ($\rho=1/12,\,1/3$) are shown. Comparing with the
phase diagram at the low filling factor, the ones at intermediate
fillings assume a more delicate Mott-insulator to self-bound superfluid
transition boundary. 

Taking the phase diagram of the system at $\rho=1/12$ for instance
{[}see Fig.~\ref{fig:intermediate_filling}(a){]}, one can notice
that in the relative small hopping regime, by increasing the attractive
interaction strength, the system first transits from self-bound superfluid
to Mott-insulator, similar to what happens in the above low filling
case, however, it transits back to self-bound superfluid from Mott-insulator
upon further increasing the attractive interaction strength {[}see
the vertical arrow and inset in Fig.~\ref{fig:intermediate_filling}(a){]}.
This manifests the attractive interaction can drive a reentrant transition
to the self-bound superfluid phase. By comparing the density and superfluid
order parameter distributions of the self-bound superfluid phase at
weak attractive interaction strength {[}see the plots with the label
1 in Fig.~\ref{fig:intermediate_filling}(b){]} and the ones of the
reentered self-bound superfluid phase {[}see the plots with the label
3 in Fig.~\ref{fig:intermediate_filling}(b){]}, one notice that
the density and superfluid order parameter distributions of the reentered
self-bound superfluid are much more compressed due to the stronger
attractive interaction. Moreover, further comparing the density distribution
of the reentered self-bound superfluid with the ones of the Mott-insulator
phases nearby in the parameter space {[}see the plots with the labels
2 and 4 in Fig.~\ref{fig:intermediate_filling}(b){]}, one can notice
that the extent of compression of the reentered self-bound superfluid
is in between the ones of these two Mott-insulators. This suggests
the reentered self-bound superfluid phase can be regarded as the intermediate
phase between two adjacent (in the parameter space) Mott-insulators
with different density distributions. Indeed, in the parameter regime
where two adjacent Mott-insulators assume similar energy, one expects
that the quantum tunneling of polarons between different sites becomes
much easier, hence gives rise to the reentered self-bound superfluid.

As a matter of fact, at larger filling factor ($\rho=1/3$ for instance),
the attractive interaction can drive not only one but a series of
reentrance to the self-bound superfluid as shown in Fig.~\ref{fig:intermediate_filling}(c).
This series of reentered self-bound superfluid appears as intermediate
phases between a series of adjacent (in the parameter space) Mott-insulators
with different density distributions {[}see Fig.~\ref{fig:intermediate_filling}(d){]}.

\begin{figure}[h]
\begin{centering}
\includegraphics[width=3.2in]{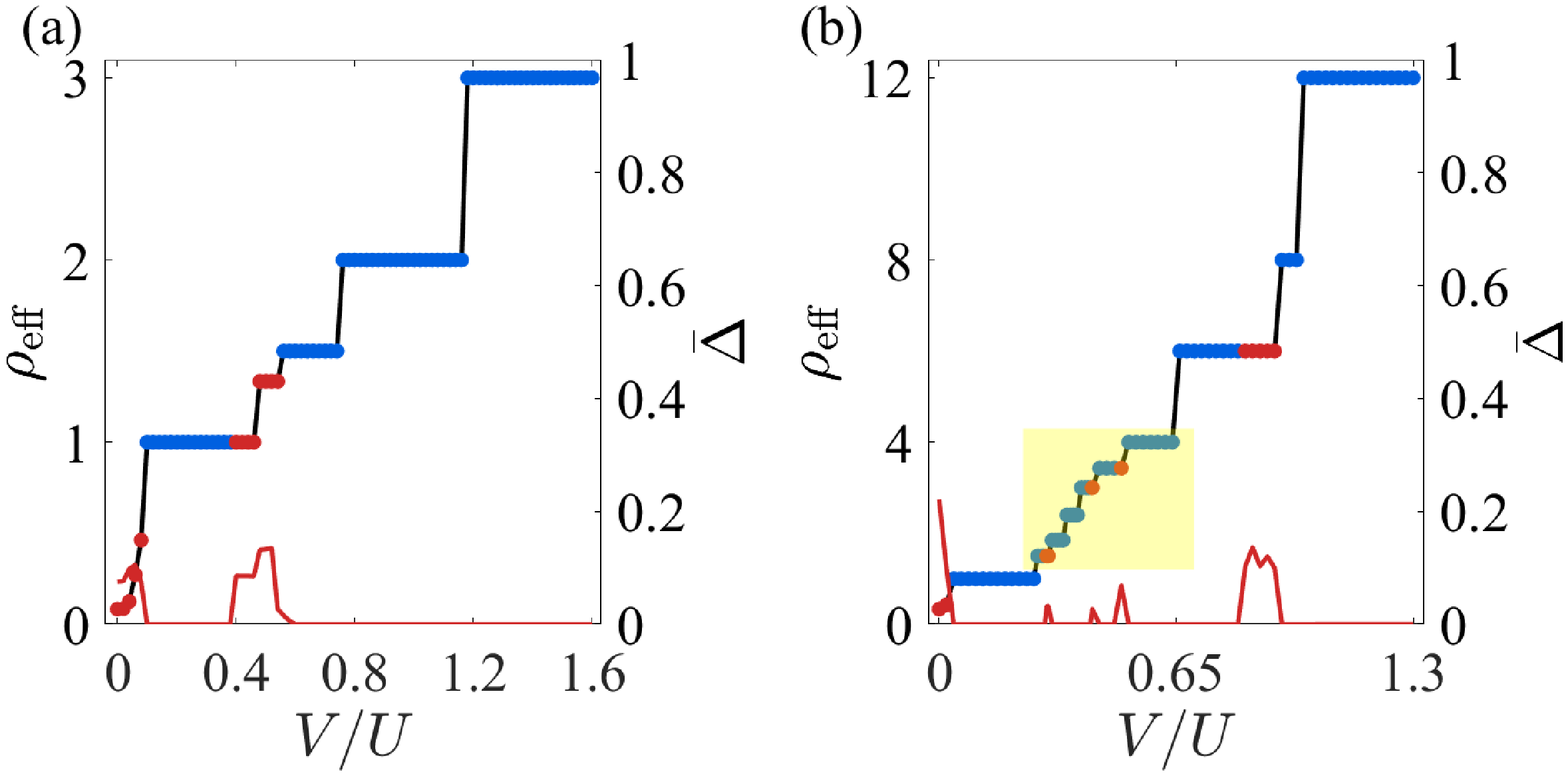}
\par\end{centering}
\begin{centering}
\includegraphics[width=3.2in]{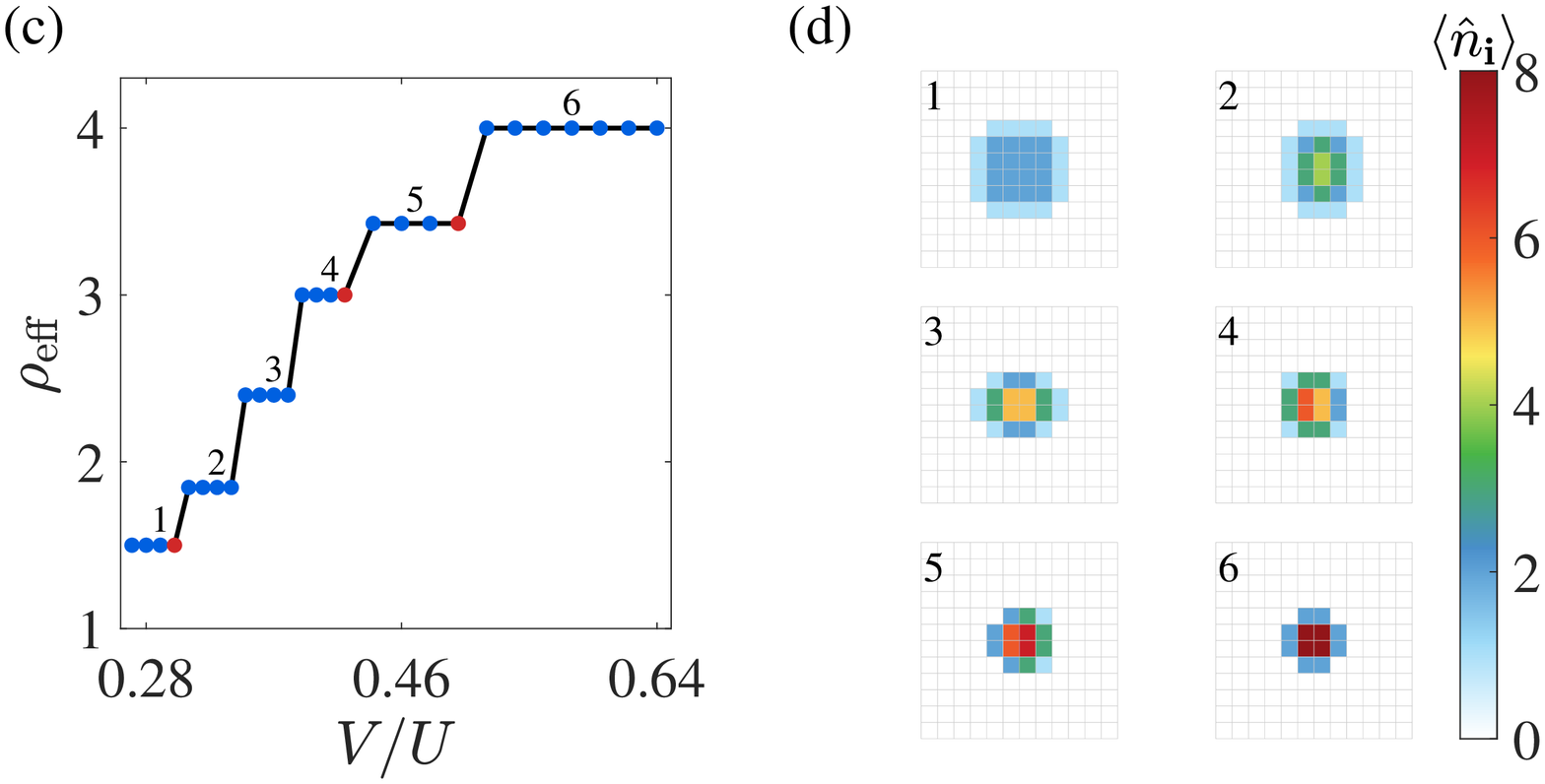}
\par\end{centering}
\caption{\label{fig:devil_staircase}Devil's staircase induced by the attractive
interaction at fixed filling factors. (a) Devil's staircase at a fixed
filling factor $\rho=1/12$. At a hopping amplitude ($2JZ/U=0.13$),
the $V/U$ dependence of the effective density $\rho_{\mathrm{eff}}$
(blue and red dots) manifests a series of plateaus. In particular,
in the large part of these plateaus marked in blue, the system is
in the incompressible Mott-insulator state as shown by the red solid
curve, which is the $V/U$ dependence of the spatial averaged local
density fluctuation $\bar{\Delta}\equiv N_{\mathrm{lat}}^{-1}\sum_{\mathbf{i}}\Delta_{\mathbf{i}}$.
(b) Devil's staircase at the filling factor $\rho=1/3$ ($2JZ/U=0.04$).
(c) Devil's staircase after zooming in the yellow region in (b). (d)
Density distributions that correspond to the blue part of each plateau
in (c). See text for more details.}
\end{figure}

To effectively characterize this series of Mott-insulators and reentered
self-bound superfluid in the weak hopping regime, we introduce the
effective density $\rho_{\mathrm{eff}}\equiv N/N_{\mathrm{lat}}^{\mathrm{eff}}$
which describes the average density of the system in the region with
nonzero polaron density ($N_{\mathrm{lat}}^{\mathrm{eff}}$ is the
number of lattice sites with nonzero density). Fig.~\ref{fig:devil_staircase}
shows how the effective density changes with respect to attractive
interaction strength at two fixed filling factors with $\rho=1/12$
and $\rho=1/3$. We notice that the effective density $\rho_{\mathrm{eff}}$
of the system manifests a series of plateaus with respect to the attractive
interaction strength, and in the large part of these plateaus marked
in blue in Fig.~\ref{fig:devil_staircase}, the system is in the
incompressible Mott-insulator state. In particular, at relatively
high filling ($\rho=1/3$ for instance) shown in Figs.~\ref{fig:devil_staircase}(b,~c),
these plateaus can be quite dense in the parameter space. This is
reminiscent of the devil's staircase in systems with long-range repulsive
interactions \citep{PhysRevB.Hubbard.1978,PhysRevLett.Fisher.1980,PhysRevB.Bak.1980.devil,PhysRevLett.Bak.1982.devil,PhysRevB.Burnell.2009.devil,PhysRevLett.Capogrosso.2010,PhysRevA.Ohgoe.2012.devil,PhysRevLett.Lan.2015,PhysRevB.Lan.2018,PhysRevA.Zhang.2021.dipolar.devil},
therefore we also refer to this succession of incompressible ground
states, dense in the parameter space, as the devil's staircase. 

However, we emphasize that there are substantial differences between
the devil's staircase found here and the ones in systems with long-range
repulsive interactions \citep{PhysRevB.Hubbard.1978,PhysRevLett.Fisher.1980,PhysRevB.Bak.1980.devil,PhysRevLett.Bak.1982.devil,PhysRevB.Burnell.2009.devil,PhysRevLett.Capogrosso.2010,PhysRevA.Ohgoe.2012.devil,PhysRevLett.Lan.2015,PhysRevB.Lan.2018,PhysRevA.Zhang.2021.dipolar.devil}.
In the latter, the devil's staircase is driven by changing the chemical
potential (or equivalently the amount of particle in the system),
i.e., different incompressible ground state that locates on each different
step of the staircase corresponds to the system with a different particle
number (or filling factor). While for the many-body polaron system
investigated here, the devil's staircase is driven by the attractive
interaction with the number of particles in the system kept fixed,
i.e., different incompressible ground state that locates on each different
step of the staircase corresponds to the system with different attractive
interaction strength but with the same particle number. Noticing also
that for systems with long-range repulsive interactions, the long
interaction range (i.e., the strength of the interaction assuming
a power law decay with respect to the distance) is crucial to give
rise to the chemical-potential-driven devil's staircase \citep{PhysRevB.Burnell.2009.devil,PhysRevLett.Capogrosso.2010,PhysRevA.Ohgoe.2012.devil,PhysRevA.Zhang.2021.dipolar.devil},
while for the many-body polaron system investigated here, the interaction
that drives the emergence of the devil's staircase is essentially
short-ranged, since the interaction range only covers nearest-neighbor
sites as shown in Hamiltonian (\ref{eq:Hamiltonian}). Moreover, the
incompressible ground states associated with the devil's staircase
in these two cases also manifest distinct spatial structures. For
systems with long-range repulsive interactions, the density distributions
of these states usually assume density wave structures commensurate
with the underlying lattice \citep{PhysRevB.Burnell.2009.devil,PhysRevLett.Capogrosso.2010,PhysRevA.Ohgoe.2012.devil,PhysRevA.Zhang.2021.dipolar.devil},
which is in sharp contrast to the self-concentrated structure in the
many-body polaron system {[}see Fig.~\ref{fig:devil_staircase}(d)
for instance{]}. 

\subsection{Experimental observability\label{subsec:Experimental-observability} }

We expect the physics predicted in this work can be readily observed
in current experimental setups. For instance, one could employ an
experimental setup similar to the one presented in Ref.~\citep{PhysRevLett.Lukas.2017.CsRb}.
Namely, one could immerse $^{133}\mathrm{Cs}$ impurities, with their
scattering lengths being $220a_{0}$ ($a_{0}$ denotes the Bohr radius),
trapped by laser beams with wavelength $\lambda=1064\mathrm{nm}$
in a BEC of an average density $n_{0}=1.0\times10^{14}\mathrm{cm^{-3}}$
formed by $^{87}\mathrm{Rb}$ atoms. By using the Feshbach resonance
between the $^{133}\mathrm{Cs}$ and $^{87}\mathrm{Rb}$ \citep{PhysRevA.Takekoshi.2012.CsRbscattering},
as shown in Fig.~\ref{fig:Estimation} in Appendix~\ref{App:Exp_Par_Est},
one can indeed tune the ratio between the attractive interaction strength
and the on-site repulsive interaction strength, i.e., $V/U$ in the
interval $(0,1.5)$ (see Appendix~\ref{App:Exp_Par_Est} for estimation
details). Moreover, one could also immerse $\mathrm{^{39}K}$ impurities
(with their scattering lengths being $278a_{0}$ \citep{PhysRevA.DalgarnoA.1998.Kscattering})
in a BEC of an average density $n_{0}=2.3\times10^{14}\mathrm{cm^{-3}}$
formed by $^{87}\mathrm{Rb}$ atoms \citep{PhysRevLett.Nils.2016.bosepolaronexper}.
Similarly, by using the Feshbach resonance between the $^{39}\mathrm{K}$
and $^{87}\mathrm{Rb}$ \citep{PhysRevLett.Nils.2016.bosepolaronexper},
as shown in Fig.~\ref{fig:Estimation} in Appendix~\ref{App:Exp_Par_Est},
one can tune the interaction ratio $V/U$ in the interval $(0,1.5)$
to observe the physics predicted here. 

Moreover, we expect that the physics predicted here could be relevant
for quantum gases consisting of atoms with magnetic dipole moments
in square optical lattices, with the attractive interaction in Hamiltonian
(\ref{eq:Hamiltonian}) realized by imposing a magnetic field rotating
along a cone centered around the direction perpendicular to the lattice
plane \citep{PhysRevLett.Giovanazzi.2002,PhysRevLett.Yi.2007,PhysRevLett.Tang.2018.dipolar}.
Also, Hamiltonian (\ref{eq:Hamiltonian}) is expected to be relevant
for microwave dressed polar molecules in square optical lattices,
where, in particular, the attractive interaction in (\ref{eq:Hamiltonian})
can be realized by a rotating electric field with the rotating axis
perpendicular to the lattice plane \citep{Nature.Schindewolf.2022}. 

\section{Conclusions}

The competition between the local repulsive interaction and the nearest-neighbor
attractive interaction in many-body polarons in optical lattices formed
by ultracold atoms can give rise to rich physics, as its phase diagrams
at different filling factors have shown: at relatively large hopping
amplitude, the attractive interaction can drive the system being in
a self-bound superfluid phase with its particle density distribution
manifesting a self-concentrated structure. In the relatively small
hopping amplitude regime, the attractive interaction can drive the
system forming the Mott-insulator phase even though the global filling
factor is not integer. Interestingly, in the Mott-insulator regime,
the system can support a series of different incompressible Mott-insulators
with their local effective filling factors manifesting a devil's staircase
structure with respect to the strength of attractive interaction.
Detailed estimation on relevant experimental parameters shows that
these rich physics can be readily observed in current experimental
setups \citep{PhysRevLett.Nils.2016.bosepolaronexper,PhysRevLett.Lukas.2017.CsRb}.
We believe our work will stimulate both further theoretical and experimental
efforts in revealing rich physics of many-body polaron systems.
\begin{acknowledgments}
We thank Jiarui Fang and Tao Yin for helpful discussions. This work
was supported by NSFC (Grant~Nos.~11874017, 12135018, 12047503,
and 12275089), NKRDPC (Grant~Nos.~2021YFA0718304 and 2022YFA1405304),
Guangdong Provincial Key Laboratory (Grant~No.~2020B1212060066),
and START Grant of South China Normal University.
\end{acknowledgments}

\appendix

\section{Derivation of the effective Hamiltonian and experimental parameter
estimation\label{App:Exp_Par_Est}}

In this appendix, we present the detailed derivation of the effective
Hamiltonian (\ref{eq:Hamiltonian}) and estimate the region of relevant
experimental parameters where the physics predicted in this work can
be observed. We consider impurities with mass $m_{I}$ interacting
with a BEC formed by atoms with mass $m_{B}$. The impurities are
trapped in a relatively deep optical lattice and described by the
Hamiltonian $\hat{H}_{I}$ presented in the main text. In the following,
we use the harmonic approximation for the Wannier basis (at site $\mathbf{0}$
for instance) for the impurity, which assumes the form
\begin{equation}
W(\mathbf{r})=\frac{1}{(\pi\sigma_{\|}^{2})^{\frac{1}{2}}}e^{-\frac{(x^{2}+y^{2})}{2\sigma_{\|}^{2}}}\left(\frac{\sqrt{n_{\perp}/n_{\|}}}{\pi\sigma_{\|}^{2}}\right)^{\frac{1}{4}}e^{-\frac{\sqrt{n_{\perp}/n_{\|}}}{2\sigma_{\|}^{2}}z^{2}},
\end{equation}
with $\sigma_{\|}=\sqrt{\hbar/m_{I}\omega_{\|}}=d/(\pi n_{\|}^{\frac{1}{4}})$
and oscillation frequency $\hbar\omega_{\|}\equiv2(V_{I}^{\|}E_{R})^{1/2}$.
The trapping strength in the transverse ($z$) direction $V_{I}^{\perp}$
is much stronger than in the parallel ($x,y$) directions $V_{I}^{\|}$
with $n_{\perp}\equiv V_{I}^{\perp}/E_{R}$ and $n_{\|}\equiv V_{I}^{\|}/E_{R}$.
The on-site interaction and hopping amplitude for the impurities can
be obtained, i.e., 
\begin{align}
U_{0} & =\frac{g_{\mathrm{II}}}{2}\int d^{3}\mathbf{r}|W(\mathbf{r})|^{4}\\
 & \approx\sqrt{8\pi}\frac{a_{\mathrm{II}}}{d}n_{\|}^{\frac{1}{2}}n_{\perp}^{\frac{1}{4}}E_{R},\nonumber 
\end{align}
\begin{equation}
J_{0}\approx\frac{4}{\sqrt{\pi}}E_{R}n_{\|}^{\frac{3}{4}}e^{-2\sqrt{n_{\|}}},
\end{equation}
where $d=\lambda/2$ is the lattice constant, $\lambda$ is the laser
wavelength and $E_{R}\equiv\hbar^{2}k^{2}/(2m_{I})$ is the recoil
energy with $k=2\pi/\lambda$. Interaction between impurities is determined
by $g_{\mathrm{II}}=4\pi\hbar^{2}a_{\mathrm{II}}/m_{I}$, where $a_{\mathrm{II}}$
is the scattering length between impurities.

The BEC of ultracold atoms with weak repulsive contact interactions
can be described by the Bogoliubov theory and treated as a phonon
bath \citep{PhysRevA.Bruderer.2007.Bogoliubov,NewJournalofPhysics.Bruderer.2008.Bogoliubov,PhysRevA.Yin.2015}
described by the Hamiltonian $\hat{H}_{B}$ presented in the main
text. The spectrum $\hbar\omega_{\mathbf{q}}$ for the Bogoliubov
phonons that appears in $\hat{H}_{B}$ assumes the explicit form $\hbar\omega_{\mathbf{q}}=\sqrt{\epsilon_{\mathbf{q}}(\epsilon_{\mathbf{q}}+2g_{\mathrm{BB}}n_{0})}$
with $\epsilon_{\mathbf{q}}\equiv\hbar^{2}|\mathbf{q}|^{2}/(2m_{B})$,
$n_{0}$ being the average BEC density, and $g_{\mathrm{BB}}$ being
the strength of the repulsive contact interaction determined by the
boson-boson scattering length $a_{\mathrm{BB}}$ via $g_{\mathrm{BB}}=4\pi\hbar^{2}a_{\mathrm{BB}}/m_{B}$. 

The impurity-BEC interaction term can be written as a Fr\"ohlich
impurity-phonon coupling \citep{PhysRevA.Yin.2015} $\hat{H}_{\mathrm{int}}$
presented in the main text, where the explicit form of $M_{\mathbf{q}}$
that appears in $\hat{H}_{\mathrm{int}}$ reads
\begin{equation}
M_{\mathbf{q}}=g_{\mathrm{IB}}\sqrt{\frac{n_{0}\epsilon_{\mathbf{q}}}{\Omega(\hbar\omega_{\mathbf{q}})^{3}}}e^{-\frac{(q_{x}^{2}+q_{y}^{2})\sigma_{\|}^{2}+q_{z}^{2}\sigma_{\perp}^{2}}{4}},
\end{equation}
with $\varOmega$ being the system quantization volume. The inter-species
interaction $g_{\mathrm{IB}}$ is determined by $g_{\mathrm{IB}}=2\pi\hbar^{2}a_{\mathrm{IB}}/m_{\mathrm{IB}}$
with $m_{\mathrm{IB}}=m_{I}m_{B}/(m_{I}+m_{B})$ being the reduced
mass and $a_{\mathrm{IB}}$ being the impurity-boson scattering length. 

As presented in the main text, one can use the Lang-Firsov polaron
transformation \citep{PhysRevA.Bruderer.2007.Bogoliubov,NewJournalofPhysics.Bruderer.2008.Bogoliubov,PhysRevB.Maier.2011.Lang_Firsov,PhysRevA.Yin.2015}
to transform the Hamitonian of the whole system $\hat{H}_{\mathrm{sys}}$
into a Hamiltonian $\tilde{H}$. This transformed Hamiltonian $\widetilde{H}$
can be separated into a coherent part $\langle\widetilde{H}\rangle$
and an incoherent part. The incoherent part is strongly suppressed
at low temperature regime $k_{B}T\ll g_{\mathrm{IB}}^{2}/(2\xi g_{\mathrm{BB}})$
\citep{PhysRevA.Bruderer.2007.Bogoliubov} with $\xi$ being the condensate
healing length. Therefore, for investigating the ground state properties
of the system, one can neglect the incoherent part and focus on the
coherent one, which is decoupled from the phonon bath and assumes
the form of an extended (polaronic) Hubbard model with phonons eliminated
by thermal averaging, i.e., 
\begin{align}
\hat{H}_{\mathrm{P}}\equiv\langle\widetilde{H}\rangle= & -\sum_{\langle\mathbf{i},\mathbf{j}\rangle}J\hat{b}_{\mathbf{i}}^{\dagger}\hat{b}_{\mathbf{j}}-\sum_{\mathbf{i}}\mu\hat{n}_{\mathbf{i}}\nonumber \\
 & +\sum_{\mathbf{i}}\frac{U_{0}-V_{\mathbf{i},\mathbf{i}}}{2}\hat{n}_{\mathbf{i}}(\hat{n}_{\mathbf{i}}-1)\\
{\color{blue}} & -\sum_{\mathbf{i}\neq\mathbf{j}}\frac{V_{\mathbf{i},\mathbf{j}}}{2}\hat{n}_{\mathbf{i}}\hat{n}_{\mathbf{j}}.\nonumber 
\end{align}
Here, $J$ is the renormalized polaronic hopping with $J\equiv J_{0}e^{-\sum_{\mathbf{q}}(2N_{\mathbf{q}}+1)[1-\cos(\mathbf{q}\cdot\mathbf{d})]|\lambda_{\mathbf{q}}M_{\mathbf{q}}|^{2}}$,
and $\mathbf{d}$ being $d\vec{e}_{x}$ or $d\vec{e}_{y}.$ $\mu$
is the renormalized chemical potential with $\mu\equiv\mu_{0}+\sum_{\mathbf{q}}\omega_{\mathbf{q}}\lambda_{\mathbf{q}}\left(2-\lambda_{\mathbf{q}}\right)\left|M_{\mathbf{q}}\right|^{2}$,
and $U_{0}-V_{\mathbf{i},\mathbf{i}}$ is the on-site interaction
strength including the polaron energy shift. The effective off-site
interaction strength
\begin{equation}
V_{\mathbf{i},\mathbf{j}}=\sum_{\mathbf{q}}\hbar\omega_{\mathbf{q}}M_{\mathbf{q}}^{2}[(2\lambda_{\mathbf{q}}-\lambda_{\mathbf{q}}^{2})+\text{h.c.}]\cos(\mathbf{q}\cdot\mathbf{R}_{\mathbf{i}\mathbf{j}}),
\end{equation}
where $\mathbf{R}_{\mathbf{i}\mathbf{j}}\equiv\mathbf{r}_{\mathbf{i}}-\mathbf{r}_{\mathbf{j}}$.
Actually, the strength of $V_{\mathbf{i},\mathbf{j}}$ decays very
fast with respect to $|\mathbf{R}_{\mathbf{i}\mathbf{j}}|$ \citep{PhysRevA.Yin.2015},
therefore, we only keep the nearest-neighbor interaction term in the
final effective Hamiltonian (\ref{eq:Hamiltonian}) for the polarons.
Moreover, for estimating the value of the parameters appearing in
the effective Hamiltonian (\ref{eq:Hamiltonian}) of the polarons,
we further employ a simple momentum independent ansatz for $\lambda_{\mathbf{q}}$,
i.e., $\lambda=\lambda_{\mathbf{q}}$. This momentum independent ansatz
works well in the strong coupling regime \citep{PhysRevA.Bruderer.2007.Bogoliubov}
and also there have been investigations showing that the variation
of $\lambda_{\mathbf{q}}$ with respect to $\mathbf{q}$ is usually
small \citep{PhysRevA.Yin.2015}, therefore we expect this ansatz
could give reasonably good estimation on the parameters that appear
in the effective Hamiltonian of the polarons, particularly in the
strong impurity-photon coupling regime that accommodates more interesting
physics. In practice, $\lambda$ is determined by minimizing the ground-state
energy \citep{PhysRevA.Yin.2015} and the corresponding self-consistent
equation for $\lambda$ reads
\begin{equation}
\lambda=\left[1+2|J_{0}|\frac{\sum_{\mathbf{q}}f_{\mathbf{q}}|M_{\mathbf{q}}|^{2}}{\sum_{\mathbf{q}}\hbar\omega_{\mathbf{q}}|M_{\mathbf{q}}|^{2}}e^{-\lambda^{2}\sum_{\mathbf{q}}f_{\mathbf{q}}|M_{\mathbf{q}}|^{2}}\right]^{-1},
\end{equation}
where $f_{\mathbf{q}}\equiv(2N_{\mathbf{q}}+1)[1-\cos(\mathbf{q}\cdot\mathbf{d})]$
with the thermally averaged phonon occupation number $N_{\mathbf{q}}\equiv[e^{\hbar\omega_{\mathbf{q}}/(k_{B}T)}-1]^{-1}$.

\begin{figure}[h]
\begin{centering}
\includegraphics[width=3.3in]{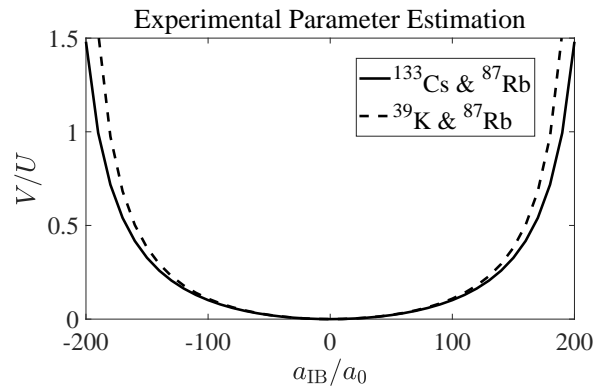}
\par\end{centering}
\caption{\label{fig:Estimation}Estimation of $V/U$ in relevant experimental
systems. Solid curve: $^{133}\mathrm{Cs}$ impurities in a BEC formed
by $^{87}\mathrm{Rb}$ atoms. Dotted curve: $^{39}\mathrm{K}$ impurities
in a BEC formed by $^{87}\mathrm{Rb}$ atoms. By tuning the scattering
length between impurity atoms and BEC atoms using Feshbach resonance,
the parameter region of $V/U$ in the phase diagrams shown in Fig.~\ref{fig:low_filling}
and Fig.~\ref{fig:intermediate_filling} can be achieved in these
experimental setups.}
\end{figure}

According to the above expressions for the interaction parameters
that appear in the effective Hamiltonian (\ref{eq:Hamiltonian}),
we can estimate the ratio between the attractive interaction strength
and the on-site repulsive interaction strength, i.e., $V/U$. The
dependence of this ratio on impurity-boson scattering length $a_{\mathrm{IB}}$
is shown in Fig.~\ref{fig:Estimation} for two relevant experimental
setups (see Sec.~\ref{subsec:Experimental-observability}). One can
see that the parameter region of $V/U$ that accommodates the physics
predicted here can be achieved by tuning $a_{\mathrm{IB}}$ in experiments.

\end{document}